\documentclass[twocolumn,showpacs,showkeys,preprintnumbers,amsmath,amssymb,superscriptaddress]{revtex4}

\usepackage{graphicx}
\usepackage{dcolumn}
\usepackage{bm}

\def\bea{\begin{eqnarray}}
\def\eea{\end{eqnarray}}

\begin{document}

\preprint{Version 3.3}

\title{The energy dependence of $p_t$ angular correlations inferred from mean-$p_{t}$ fluctuation scale dependence in heavy ion collisions at the SPS and RHIC}

\affiliation{Argonne National Laboratory, Argonne, Illinois 60439}
\affiliation{University of Birmingham, Birmingham, United Kingdom}
\affiliation{Brookhaven National Laboratory, Upton, New York 11973}
\affiliation{California Institute of Technology, Pasadena, California 91125}
\affiliation{University of California, Berkeley, California 94720}
\affiliation{University of California, Davis, California 95616}
\affiliation{University of California, Los Angeles, California 90095}
\affiliation{Carnegie Mellon University, Pittsburgh, Pennsylvania 15213}
\affiliation{Creighton University, Omaha, Nebraska 68178}
\affiliation{Nuclear Physics Institute AS CR, 250 68 \v{R}e\v{z}/Prague, Czech Republic}
\affiliation{Laboratory for High Energy (JINR), Dubna, Russia}
\affiliation{Particle Physics Laboratory (JINR), Dubna, Russia}
\affiliation{University of Frankfurt, Frankfurt, Germany}
\affiliation{Institute of Physics, Bhubaneswar 751005, India}
\affiliation{Indian Institute of Technology, Mumbai, India}
\affiliation{Indiana University, Bloomington, Indiana 47408}
\affiliation{Institut de Recherches Subatomiques, Strasbourg, France}
\affiliation{University of Jammu, Jammu 180001, India}
\affiliation{Kent State University, Kent, Ohio 44242}
\affiliation{Institute of Modern Physics, Lanzhou, China}
\affiliation{Lawrence Berkeley National Laboratory, Berkeley, California 94720}
\affiliation{Massachusetts Institute of Technology, Cambridge, MA 02139-4307}
\affiliation{Max-Planck-Institut f\"ur Physik, Munich, Germany}
\affiliation{Michigan State University, East Lansing, Michigan 48824}
\affiliation{Moscow Engineering Physics Institute, Moscow Russia}
\affiliation{City College of New York, New York City, New York 10031}
\affiliation{NIKHEF and Utrecht University, Amsterdam, The Netherlands}
\affiliation{Ohio State University, Columbus, Ohio 43210}
\affiliation{Panjab University, Chandigarh 160014, India}
\affiliation{Pennsylvania State University, University Park, Pennsylvania 16802}
\affiliation{Institute of High Energy Physics, Protvino, Russia}
\affiliation{Purdue University, West Lafayette, Indiana 47907}
\affiliation{Pusan National University, Pusan, Republic of Korea}
\affiliation{University of Rajasthan, Jaipur 302004, India}
\affiliation{Rice University, Houston, Texas 77251}
\affiliation{Universidade de Sao Paulo, Sao Paulo, Brazil}
\affiliation{University of Science \& Technology of China, Hefei 230026, China}
\affiliation{Shanghai Institute of Applied Physics, Shanghai 201800, China}
\affiliation{SUBATECH, Nantes, France}
\affiliation{Texas A\&M University, College Station, Texas 77843}
\affiliation{University of Texas, Austin, Texas 78712}
\affiliation{Tsinghua University, Beijing 100084, China}
\affiliation{Valparaiso University, Valparaiso, Indiana 46383}
\affiliation{Variable Energy Cyclotron Centre, Kolkata 700064, India}
\affiliation{Warsaw U8niversity of Technology, Warsaw, Poland}
\affiliation{University of Washington, Seattle, Washington 98195}
\affiliation{Wayne State University, Detroit, Michigan 48201}
\affiliation{Institute of Particle Physics, CCNU (HZNU), Wuhan 430079, China}
\affiliation{Yale University, New Haven, Connecticut 06520}
\affiliation{University of Zagreb, Zagreb, HR-10002, Croatia}

\author{J.~Adams}\affiliation{University of Birmingham, Birmingham, United Kingdom}
\author{M.M.~Aggarwal}\affiliation{Panjab University, Chandigarh 160014, India}
\author{Z.~Ahammed}\affiliation{Variable Energy Cyclotron Centre, Kolkata 700064, India}
\author{J.~Amonett}\affiliation{Kent State University, Kent, Ohio 44242}
\author{B.D.~Anderson}\affiliation{Kent State University, Kent, Ohio 44242}
\author{M.~Anderson}\affiliation{University of California, Davis, California 95616}
\author{D.~Arkhipkin}\affiliation{Particle Physics Laboratory (JINR), Dubna, Russia}
\author{G.S.~Averichev}\affiliation{Laboratory for High Energy (JINR), Dubna, Russia}
\author{Y.~Bai}\affiliation{NIKHEF and Utrecht University, Amsterdam, The Netherlands}
\author{J.~Balewski}\affiliation{Indiana University, Bloomington, Indiana 47408}
% ...

\collaboration{STAR Collaboration}\noaffiliation

\date{\today}

\begin{abstract}
We present the first study of the energy dependence of $p_t$ angular correlations inferred from event-wise mean transverse momentum $\langle p_{t} \rangle$ fluctuations in heavy ion collisions. We compare our large-acceptance measurements at CM energies $\sqrt{s_{NN}} =$ 19.6, 62.4, 130 and 200 GeV to SPS measurements at 12.3 and 17.3 GeV. $p_t$ angular correlation structure suggests that the principal source of $p_t$ correlations and fluctuations is minijets (minimum-bias parton fragments). We observe a dramatic increase in correlations and fluctuations from SPS to RHIC energies, increasing linearly with  $\ln \sqrt{s_{NN}}$ from the onset of observable jet-related $\langle p_{t} \rangle$ fluctuations near 10 GeV.  \\
\end{abstract}


\maketitle

 \section{Introduction}

Theoretical descriptions of heavy ion (HI) collisions at RHIC energies predict copious parton (mainly gluon) production in the early stages of collisions and subsequent parton rescattering as the principal route to a color-deconfined bulk medium, with possible equilibration to a quark-gluon plasma~\cite{theor0,theor1,theor2,theor3,theor4}. Particle yields, spectra and high-$p_t$ correlations from Au-Au collisions at 130 and 200 GeV provide tantalizing evidence that a QCD colored medium is indeed produced at RHIC~\cite{qgp,highpt,star2,star3,brahms,phenix2}. What are the properties of that medium, and has an equilibrated QGP formed prior to hadronic decoupling? A partial answer may emerge by searching for evidence of initial-state semi-hard scattered partons in the correlations and fluctuations of final-state hadrons. In particular, $p_t$ fluctuations and correlations may provide such evidence. 

Fluctuations (variations about a mean) of event-wise mean transverse momentum $\langle p_t \rangle$~\cite{meanptprl,phenix} within momentum-space angular bins of varying size, and corresponding two-particle $p_t$ correlations (variation of a two-particle distribution relative to a reference), could provide access to early parton scattering and subsequent in-medium dissipation inaccessible by other means~\cite{flowarg}. $p_t$ correlations may represent such partons in the form of local velocity and/or temperature correlations~\cite{QT,veltemp}. Measurements of $\langle p_t \rangle$ fluctuations in Au-Au collisions at fixed scale (bin size) at 130 GeV~\cite{meanptprl} and measurements of $p_t$ angular correlations inferred from $\langle p_t \rangle$ fluctuation scale dependence at 200 GeV~\cite{ptscale} indicate that $p_t$ correlations at RHIC are much larger than those at the SPS.  

In this paper we report the first study of the energy dependence of $p_t$ angular correlations ({\em e.g.,} structures in the event-wise $p_t$ distribution on ($\eta,\phi$) which occur at different positions in each HI collision) inferred from {\em excess} $\langle p_t \rangle$ fluctuations (fluctuations beyond those expected for independent particle $p_t$ production).
We present the scale dependence of $\langle p_t \rangle$ fluctuations within the STAR detector acceptance for four RHIC energies and provide a basis for interpreting those fluctuations by inverting the fluctuation scale dependence for two energies and two centralities to form $p_t$ autocorrelations on angle space $(\eta,\phi)$. We obtain the centrality dependence of full-acceptance fluctuations at four RHIC energies compared to results at two SPS energies, and we determine  the $\sqrt{s_{NN}}$ dependence of  $\langle p_t \rangle$ fluctuations for full-acceptance STAR data as a basis for comparison with extrapolated CERES measurements~\cite{ceres} and the pQCD event simulation Monte Carlo Hijing ~\cite{hijing}. This analysis is based on Au-Au collisions observed with the STAR detector at the Relativistic Heavy Ion Collider (RHIC).

\section{Analysis Method}

Excess {\em charge-independent} (all charged particles combined) $\langle p_t \rangle$ fluctuations are measured by the difference between the variance of quantity \mbox{$\{p_t(\delta x)  - n(\delta x)\, \hat p_t \} / \sqrt{\bar n(\delta x)}$} and the variance $\sigma^2_{\hat p_t}$ of an {\em uncorrelated} reference~\cite{meanptprl}. $p_{t}(\delta x)$ is the scalar sum of $p_t$ in a bin of size $\delta x$ ({\em e.g.,} $\delta \eta$ or $\delta \phi$), $n(\delta x)$ is the number of particles in the bin, and $\hat p_t $ is the mean and $\sigma^2_{\hat p_t}$ is the variance of the single-particle $p_t$ spectrum for all accepted charged particles from all events ($\hat p_t$ and $\sigma^2_{\hat p_t}$ then represent {\em independent} particle $p_t$ production). The {\em variance difference}
\bea \label{Eq3}
\Delta\sigma^{2}_{p_t:n}(\delta x) & \equiv & 
\overline{ \left\{
 p_{t}(\delta x) - n(\delta x)\, \hat{p}_t \right\}^2} / \bar n(\delta x) - \sigma^2_{\hat p_t} \\ \nonumber
 & \equiv & 2 \sigma_{\hat{p}_t} \Delta\sigma_{p_t:n}(\delta x),
\label{Eq4}
\eea
calculated over a range of bin sizes is the scale-dependent $\langle p_t \rangle$ fluctuation measure for this analysis~\cite{ptscale}. Overlines indicate averages over all bins with size $\delta x$ in all events. 
The {\em difference factor} $\Delta\sigma^{}_{p_t:n}$ is related to the $\langle p_t \rangle$ fluctuation measure $\Phi_{p_t}$ introduced previously~\cite{phipt,na49ptfluct,na492} by $\Delta \sigma_{p_t:n} = \Phi_{p_t}\, (1 + \Phi_{p_t} /\, 2 \sigma_{\hat p_t})$.

Event-wise fluctuations in the bin contents of a binned distribution reflect changes in the structure of that distribution. Smaller bins are sensitive to more local aspects of that structure, and conversely. The scale dependence of fluctuations is therefore equivalent in some sense to a running integral of the corresponding two-particle distribution, the integration limit determined by the bin size or scale~\cite{inverse,qingjun}. The integrand is an {\em autocorrelation} which compares a distribution $f(x)$ {to itself}. An autocorrelation is effectively a {\em projection by averaging} of product distribution $f(x_1)\cdot f(x_2)$ on $(x_1,x_2)$ onto the difference variable $x_\Delta \equiv x_1 - x_2$~\cite{diffvar}. In this analysis we wish to determine the average angular correlation structure of the event-wise $p_t$ distribution on $(\eta,\phi)$: what are the aspects of that distribution which vary event-wise but which nevertheless have persistence and universality. We do so by inverting the scale dependence of excess $\langle p_t \rangle$ fluctuations (the integral) to obtain the autocorrelation of the $p_t$ distribution on $(\eta,\phi)$~\cite{inverse,qingjun,ptscale}. 

Eq.~(\ref{inverse}) below is an {\em integral equation} in discrete form which relates variance difference $\Delta \sigma_{p_t:n}^2(\delta \eta,\delta \phi)$ on pseudorapidity $\eta$ and azimuth angle $\phi$ to an autocorrelation distribution on $(\eta_\Delta,\phi_\Delta)$ ({\em e.g.,} $\eta_\Delta \equiv \eta_1 - \eta_2$)~\cite{ptscale,inverse}. The autocorrelation ({\em cf.} Fig.~\ref{fig2} for examples) compactly represents two-particle correlations on $(\eta,\phi)$ in HI collisions~\cite{axialcd}. 
The 2D discrete integral equation is 
\bea \label{inverse}
\Delta \sigma^2_{p_t:n}(m \, \epsilon_\eta, n \, \epsilon_\phi) &= & \\ \nonumber 
 4 \sum_{k,l=1}^{m,n} \epsilon_\eta \epsilon_\phi & &\hspace{-.25in} K_{mn;kl}  \,   \frac{\Delta \rho(p_t:n;k\,\epsilon_\eta, l\, \epsilon_\phi) }{ \sqrt{\rho_{ref}(n;k\,\epsilon_\eta, l\, \epsilon_\phi)}} ,
\end{eqnarray}
with kernel $K_{mn;kl} \equiv (m - {k + 1/2})/{m} \cdot (n-{l+1/2})/{n}$ and fixed {\em microbin} sizes $\epsilon_\eta$ and $\epsilon_\phi$ for the discrete integral. That equation can be inverted (solved for the integrand) to obtain autocorrelation density ratio ${\Delta \rho(p_t:n)}/{ \sqrt{\rho_{ref}(n)}}
$ [units (GeV/c)$^2$] as a {\em per-particle} $p_t$ correlation measure on $(\eta_\Delta,\phi_\Delta)$ from $\langle p_t \rangle$ fluctuation scale dependence of $\Delta \sigma^2_{p_t:n}(\delta \eta,\delta \phi)$ \cite{ptscale,inverse}. $\Delta \rho(p_t:n)$ is proportional to the average of $(p_t - n \hat p_t)$ covariances for all pairs of bins $(\eta_\Delta,\phi_\Delta)$ apart. $\sqrt{\rho_{ref}(n)}$ is the geometric mean of particle densities in those bins~\cite{ptscale}. Density ratio ${\Delta \rho(p_t:n)}/{ \sqrt{\rho_{ref}(n)}}$ is thus proportional to {\em normalized covariance} $\overline{(p_t - n \hat p_t)_a(p_t - n \hat p_t)_b}/\sqrt{\bar n_a \bar n_b}$ (averaged over certain bin combinations ($a,b$) to form an autocorrelation). The density ratio  has the form of Pearson's correlation coefficient~\cite{pearson}, but with number variances in the denominator replaced by Poisson values $\bar n_a,~ \bar n_b$. The density ratio is derived and discussed in~\cite{inverse,mitinv,lowq2,ppmeas}.

\section{Data}

Data for this analysis were obtained with the STAR detector~\cite{starnim} using a 0.5~T uniform magnetic field parallel to the beam axis. Event triggering  and charged-particle measurements with the Time Projection Chamber (TPC) are described in \cite{starnim}. Track definitions, tracking efficiencies, quality cuts and primary-particle definition are described in~\cite{meanptprl,spectra}. Tracks were accepted with pseudorapidity in the range $|\eta| < $ 1, transverse momentum in the range $p_t \in [0.15,2]$ GeV/c and $2\pi$ azimuth, defining the detector acceptance for this analysis. Particle identification was not implemented. Centrality classes (percentages of the total hadronic cross section) were defined in terms of the uncorrected number $N$ of charged particles in acceptance $|\eta|< 1$ according to procedures described in~\cite{meanptprl,ptscale}.

Centrality specified in terms of pathlength $\nu$ (estimating the mean number of nucleons encountered by a participant nucleon) is based on the relationship of $N$ to minimum-bias distribution endpoints $N_p$ and $N_0$. 
$N_p$, the lower half-maximum point of the minimum-bias distribution plotted as $d\sigma / dN^{1/4}$, estimates the uncorrected mean multiplicity for non-single-diffractive nucleon-nucleon collisions in the same acceptance. $N_0$, the upper half-maximum point of $d\sigma / dN^{1/4}$, estimates the value of $N$ corresponding to the maximum number of participant nucleons $N_{part,max}$ and impact parameter $b = 0$. The relation between the fractional cross section $\sigma/\sigma_0$ and multiplicity $N$ is approximated by expression $1 - \sigma / \sigma_0 = (N^{1/4} - N_p^{1/4}) / (N_0^{1/4} - N_p^{1/4})$~\cite{hijcent}, accurate to $\sim$ 3\% over the entire centrality range (exluding fluctuations near the endpoints). Mean participant path length $\nu \approx 2 N_{bin} / N_{part}$~\cite{nu} is then given by $\nu = \{1 + 2.23\, (1 - \sigma / \sigma_0)\}^{6}/\{(1 + 2.72\, (1 - \sigma / \sigma_0)\}^{4}$ to about 2\%, based on a Monte Carlo Glauber simulation~\cite{hijcent}. Coefficients 2.23 and 2.72 apply to 200 GeV collisions and vary slowly with $\ln(\sqrt{s_{NN}})$, resulting in a few-percent shift in $\nu_{max}$ with collision energy over the energy range 62 - 200 GeV. Those expressions determined the values of $\nu$ used in Fig.~\ref{fig3} (right panel).

\section{Fluctuations and Correlations}

\begin{figure}[t]
\begin{tabular}{cc}
\begin{minipage}{.47\linewidth}
\includegraphics[keepaspectratio,width=1.65in]{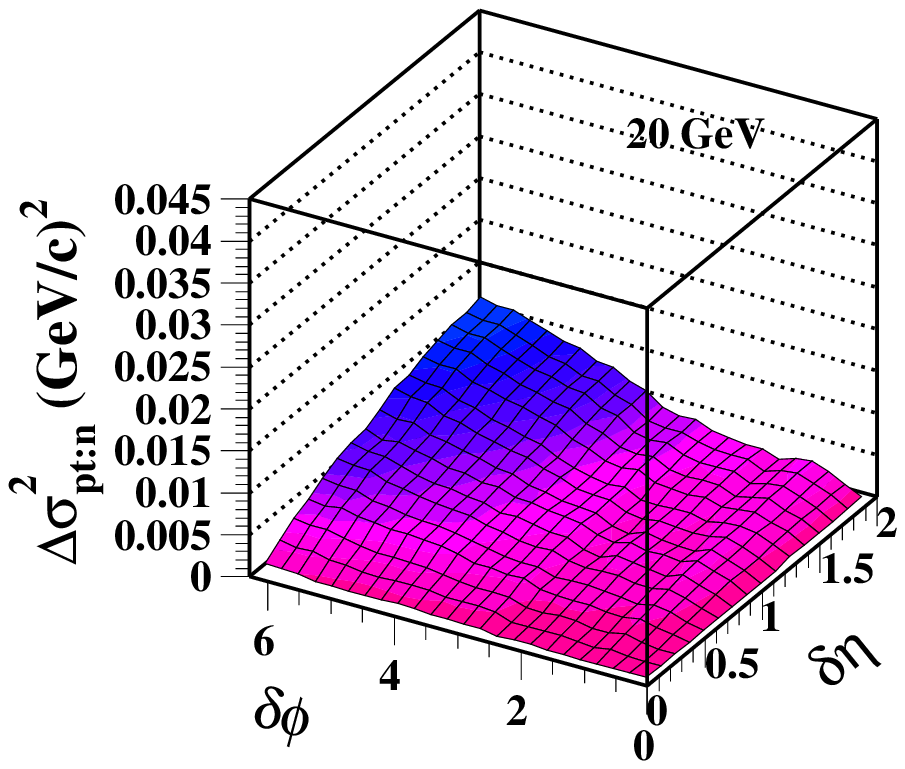}
\end{minipage} & 
\begin{minipage}{.47\linewidth}
\includegraphics[keepaspectratio,width=1.65in]{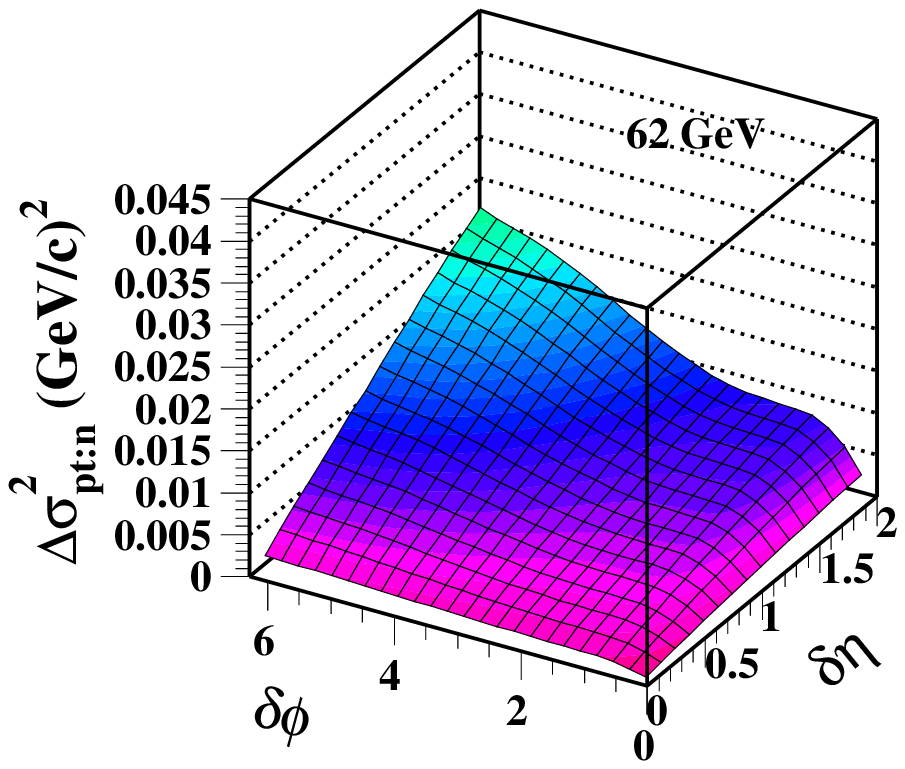}
\end{minipage} \\ 
\begin{minipage}{.47\linewidth}
\includegraphics[keepaspectratio,width=1.65in]{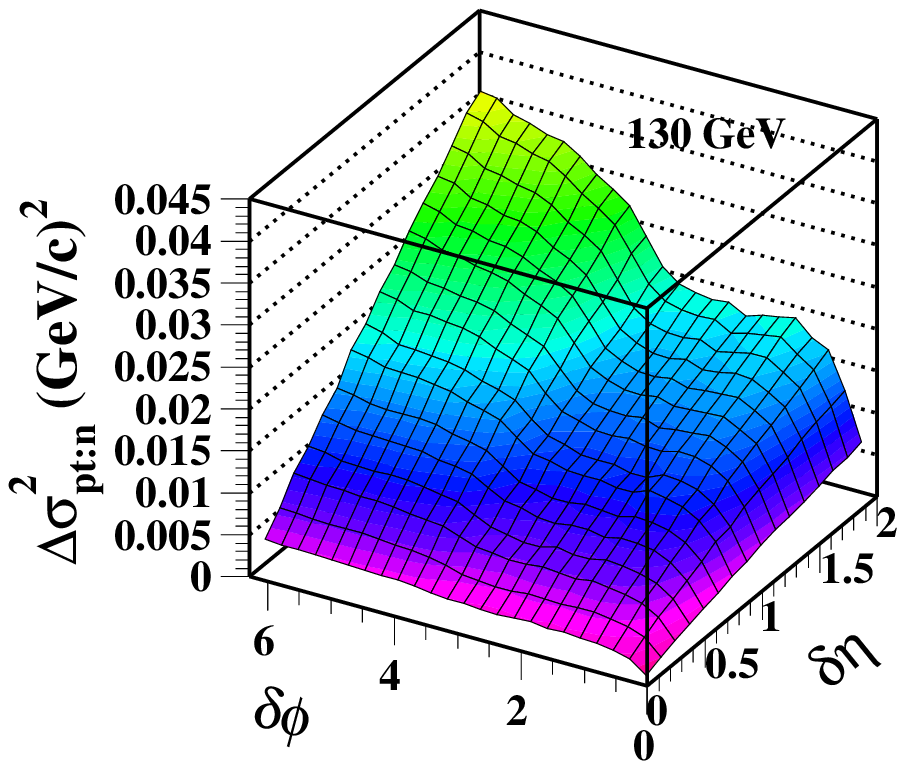}
\end{minipage}&
\begin{minipage}{.47\linewidth}
\includegraphics[keepaspectratio,width=1.65in]{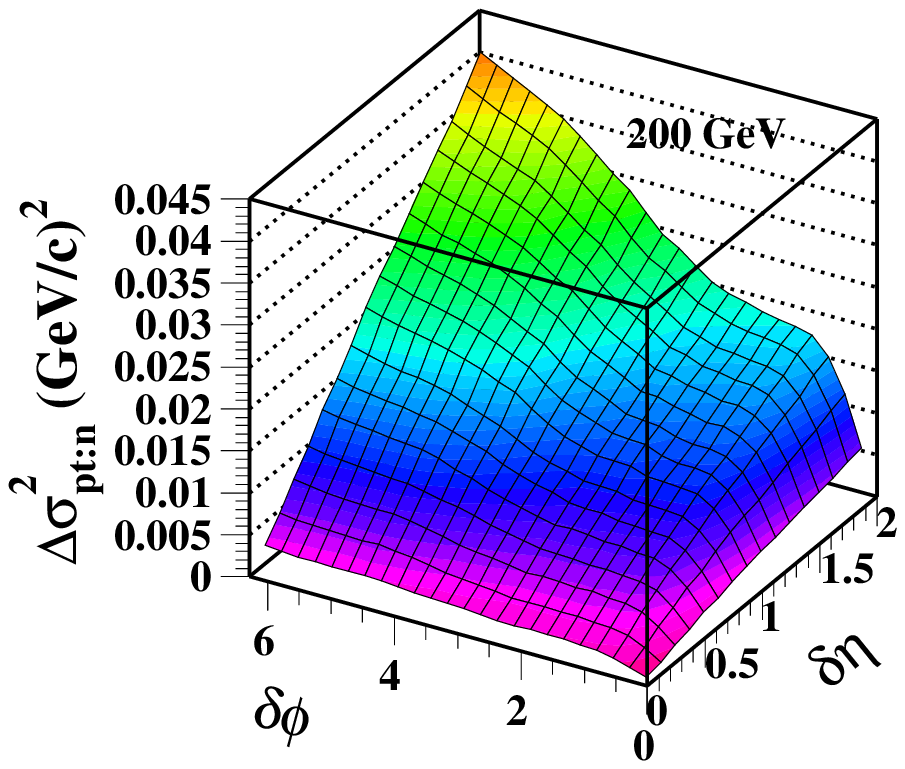}
\end{minipage}\\
\end{tabular}
\caption{Per-particle $\langle p_t \rangle$ fluctuation scale dependence for 19.6, 62.4, 130 and 200 GeV Au-Au collisions, and for the top 0-20, 0-5, 0-15 and 0-5\% of the total hadronic cross section respectively.  \label{fig1}}
\end{figure}

Fig.~\ref{fig1} compares the scale dependence of variance difference $\Delta \sigma^2_{p_t:n}(\delta \eta,\delta \phi)$ in Eq.~\ref{Eq3} for central Au-Au collisions and four collision energies: $\sqrt{s_{NN}} =$ 19.6, 62.4, 130 and 200 GeV. The increase of fluctuation amplitudes between 19.6 and 200 GeV in Fig.~\ref{fig1} is a factor four, establishing that the $\langle p_t \rangle$ variance difference is strongly energy dependent. However, $\langle p_t \rangle$ fluctuations are difficult to interpret, whereas the corresponding  $p_t$ angular autocorrelations obtained by inverting fluctuation scale dependence clearly indicate the underlying dynamics.

\begin{figure}[t]
\begin{tabular}{cc}
\begin{minipage}{.47\linewidth}
\includegraphics[keepaspectratio,width=1.65in]{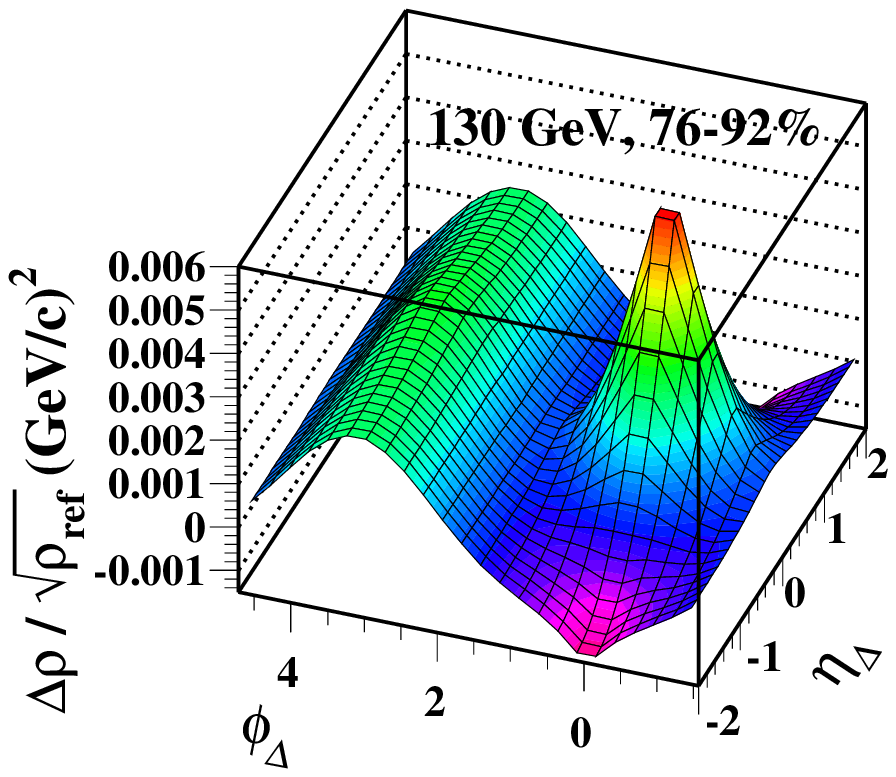}
\end{minipage}&
\begin{minipage}{.47\linewidth}
\includegraphics[keepaspectratio,width=1.65in]{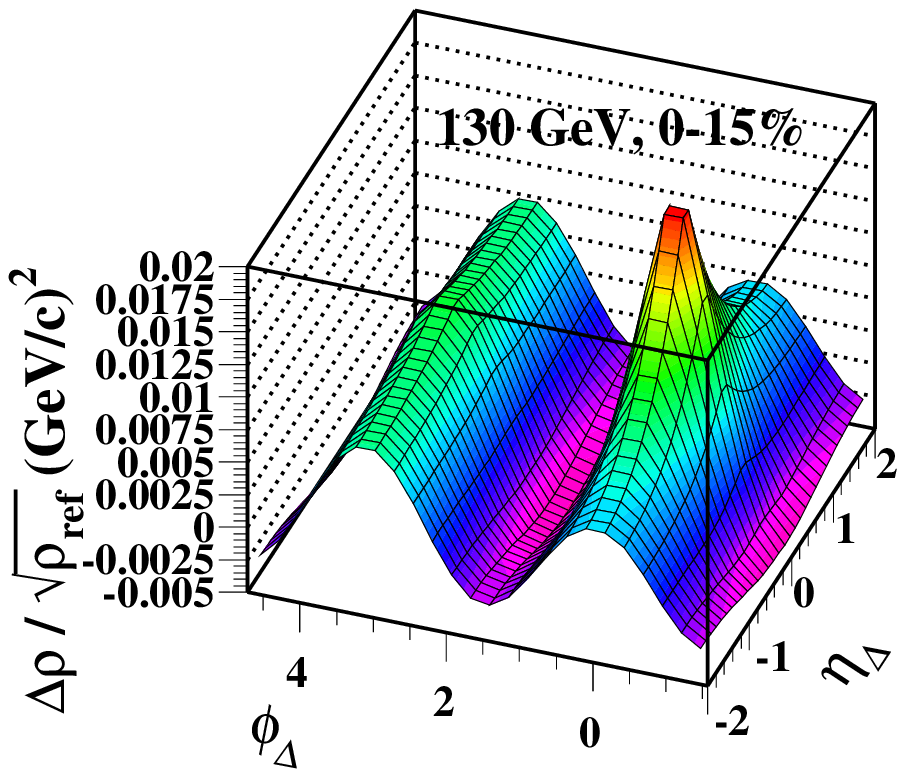}
\end{minipage}\\
\begin{minipage}{.47\linewidth}
\includegraphics[keepaspectratio,width=1.65in]{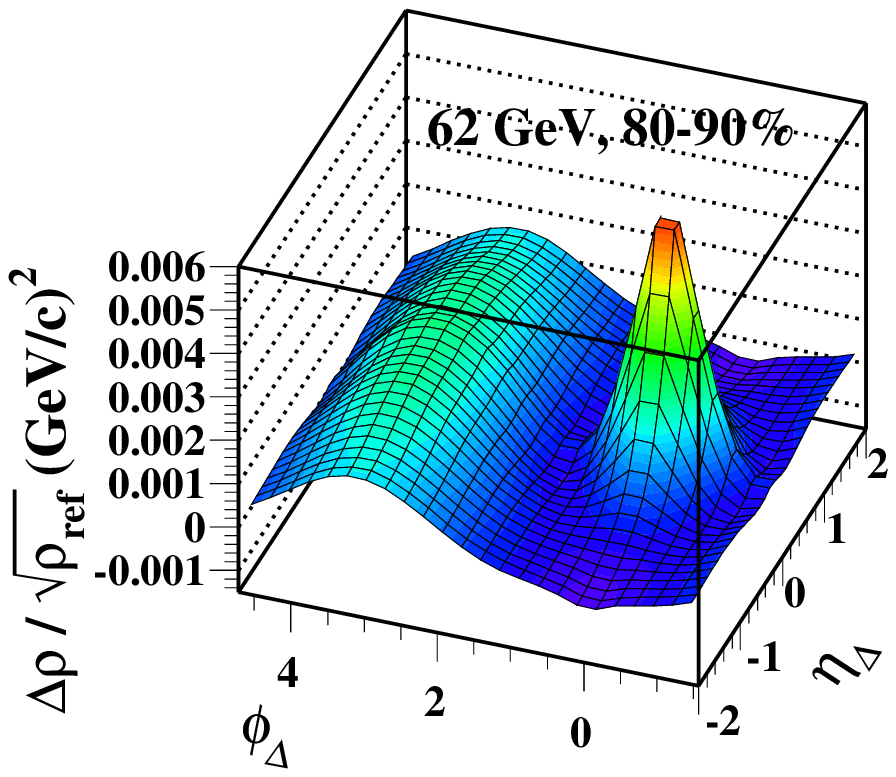}
\end{minipage} & 
\begin{minipage}{.47\linewidth}
\includegraphics[keepaspectratio,width=1.65in]{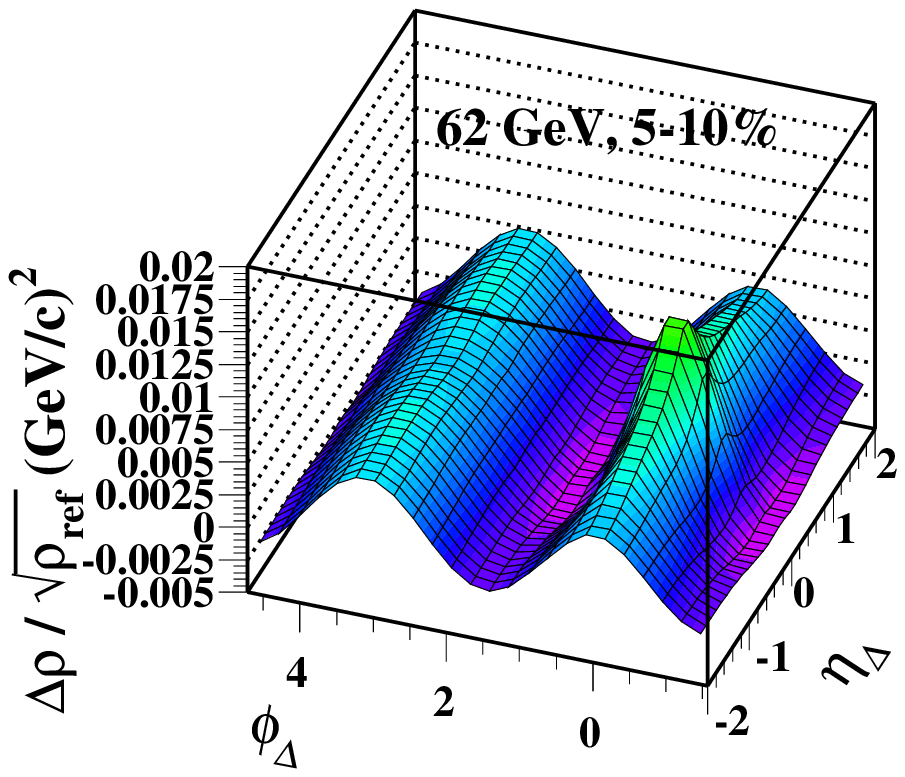}
\end{minipage}\\
\end{tabular}
\caption{$p_t$ autocorrelations for 130 GeV (upper) and 62.4 GeV (lower), and for comparable peripheral (left) and central (right) Au-Au collisions. These autocorrelations contain same-side ($|\phi_\Delta| < \pi/2$) and away-side ($|\phi_\Delta| > \pi/2$) structures.   \label{fig2}}
\end{figure}

Fig.~\ref{fig2} shows $p_t$ angular autocorrelations (by construction symmetric about $\eta_\Delta,~\phi_\Delta = 0$) of density ratio $\Delta \rho / \sqrt{\rho_{ref}}$ inferred from $\langle p_t \rangle$ fluctuation scale dependence as in Fig.~\ref{fig1} by inverting Eq.~(\ref{inverse})~\cite{inverse}. The plots in Figs.~\ref{fig1} and \ref{fig2} contain equivalent information in different forms, consistent with Eq.~(\ref{inverse}). One can observe the equivalents of the elliptic flow sinusoids in Fig.~\ref{fig2} (right panels) along the upper-right edges of the plots in Fig.~\ref{fig1}. Note that the elliptic flow sinusoid amplitudes in this autocorrelation representation using {\em per-particle} correlation measure $\Delta \rho / \sqrt{\rho_{ref}}$ are negligible for peripheral collisions and increase with increasing centrality toward a maximum for mid-central collisions. That behavior contrasts with the nearly opposite trend observed with conventional {\em per-pair} measure $v_2$. The top two panels of Fig.~\ref{fig2} represent 130 GeV and the bottom two panels 62.4 GeV Au-Au collisions. The left panels represent comparable peripheral collisions and the right panels comparable central collisions. Autocorrelations for 200 GeV are presented in~\cite{ptscale}, and the available 19.6 GeV data do not have sufficient statistics for a satisfactory fluctuation inversion. The autocorrelations in Fig.~\ref{fig2} have same-side ($|\phi_\Delta| < \pi/2$) and away-side ($|\phi_\Delta| > \pi/2$) components. 

We interpret the same-side peak and away-side ridge which dominate peripheral collisions as consistent with {minijets} as the source mechanism. The shapes are similar to jet correlations observed in p-p collisions~\cite{ppmeas}, and in Au-Au collisions of all centralities as modeled by Hijing~\cite{qingjun}. Those structures are strongly modified with increasing A-A centrality, but in a continuous manner which suggests that the minijet interpretation is also appropriate in central heavy ion collisions. By `minijet' we refer to correlated hadrons from initial-state semi-hard parton scattering and subsequent fragmentation in which no leading or trigger particle is required, {\em i.e.,} fragments from {\em minimum-bias} partons with no {\em analysis} restriction placed on the parton momentum spectrum. We expect minimum-bias partons to be dominated by the low-$Q^2$ `minijets' described by theory~\cite{theor0,theor1,theor2,theor3}. The amplitudes of the peripheral same-side peaks are notably similar for 62.4 and 130 GeV, whereas for central collisions the same-side peak amplitude increases strongly with energy, and the peak is significantly broadened in the $\eta_\Delta$ direction. 

\section{Experiment Comparisons}

The autocorrelations in Fig.~\ref{fig2} represent all the angular correlation information obtainable from corresponding fluctuation measurements. Different experimental circumstances ({\em e.g.,} detector acceptances) may result in {\em apparently} conflicting fluctuation measurements. However, {\em per-particle} fluctuation measurements such as those presented here are exactly comparable at the same bin size or scale [integration limits of Eq.~(\ref{inverse})], {\em independent of detector geometry and other experimental details}, because they integrate the underlying autocorrelations which are detector-independent distributions. $\langle p_t \rangle$ fluctuations have been measured by several collaborations~\cite{meanptprl,ptscale,na49ptfluct,na492,ceres,phenix}. PHENIX measurements at 130 and 200 GeV~\cite{phenix} are compatible with STAR measurements at equivalent acceptances (scales). We wish to determine the energy dependence of $p_t$ angular corrrelations over the largest energy interval possible, from 200 GeV at RHIC down to the lowest SPS energies. However, the CERES SPS measurements are restricted to the $\eta$ scale dependence of $\langle p_t \rangle$ fluctuations at full azimuth. We make the most differential comparisons possible with that limited information, given that fluctuation scale dependence is the running integral of the underlying autocorrelation. First, we make a detailed comparison of STAR and CERES fluctuation scale dependence at full azimuth and equivalent pseudorapidity scales. Then we examine changes in centrality dependence with energy.

\subsection{Scale dependence}

\begin{figure}[h]
\includegraphics[width=1.65in,height=1.65in]{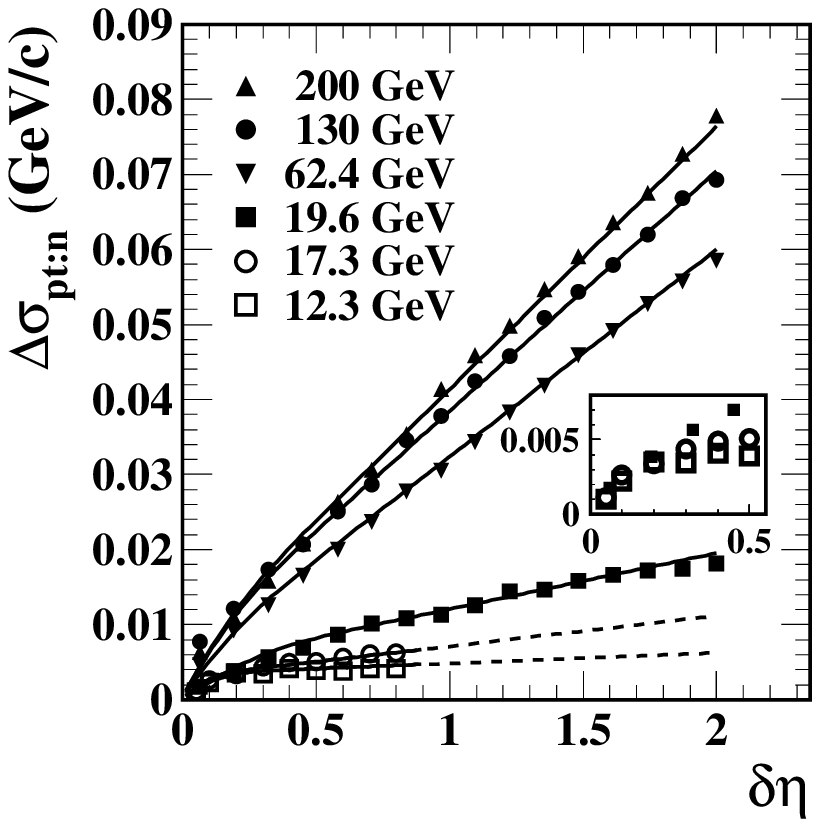}
\includegraphics[width=1.65in,height=1.65in]{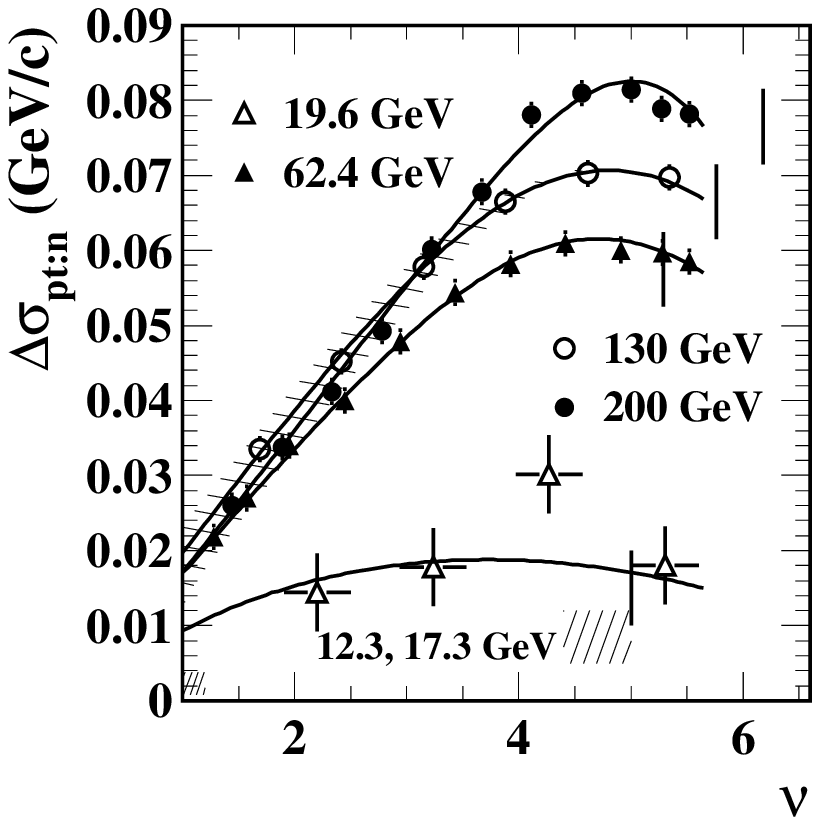}
\caption{Left panel: Per-particle fluctuation dependence on pseudorapidity scale $\delta \eta$ (in $2\pi$ azimuth) in central collisions ({\em cf.} Fig.~\ref{fig1} caption for STAR centralities). STAR measurements are solid symbols, CERES measurements~\cite{ceres} are open symbols. The inset shows details at small $\delta \eta$. Right panel: Centrality dependence of $\langle p_t \rangle$ fluctuations in the STAR acceptance for four energies. $\nu$ is the mean participant path length~\cite{nu}. The vertical line at right estimates $\nu$ for $b = 0$. The upper hatched band estimates the uncertainty in $\nu$ for 130 GeV data. There is an overall 14\% systematic error in the corrected amplitudes. SPS measurements of $\Phi_{p_t}$ at 12.3 and 17.3 GeV (the lower-right hatched region, with errors and centrality range) are included for comparison. Curves guide the eye.
\label{fig3}}
\end{figure}

Fig.~\ref{fig3} (left panel) shows CERES $\Phi_{p_t}$ values for Pb-Au collisions at several pseudorapidity scales and $2\pi$ azimuth acceptance for $\sqrt{s_{NN}} =$ 12.3 and 17.3 GeV~\cite{ceres}. Also plotted are comparable STAR measurements of $\Delta \sigma_{p_t:n}$. The CERES data rise rapidly to about 3 MeV/c within $\delta \eta \leq 0.2$ (see inset), with a slower {\em linear} rise thereafter. STAR 19.6 GeV data show similar behavior, albeit with somewhat larger magnitudes over a larger $\eta$ acceptance.  The higher-energy STAR data are qualitatively larger in magnitude. We attribute the rapid rise of CERES data ($\sim$ 0.003 GeV/c) in $\delta \eta < 0.2$ to quantum (HBT) and Coulomb correlations (resonance decays make a negligible contribution to $\langle p_t \rangle$ fluctuations) and designate those contributions as small-scale correlations (SSC). The complementary region of $\delta \eta$ then represents large-scale correlations (LSC). We conclude (see below) that the LSC component is dominated by parton fragments (including the away-side $\eta_\Delta$-independent azimuth peak), and possibly global temperature fluctuations [with corresponding autocorrelation uniform on $(\eta_\Delta,\phi_\Delta)$].  

\subsection{Centrality dependence}

Fig.~\ref{fig3} (right panel) shows the centrality dependence of charge-independent $\langle p_t \rangle$ fluctuations (elliptic flow does not contribute to $\langle p_t \rangle$ fluctuations integrated over a $2\pi$ azimuth acceptance). $\Delta \sigma_{p_t:n}$ was measured at the STAR acceptance scale and at four collision energies and corrected for tracking inefficiency and background contamination as in~\cite{meanptprl}. The vertical scale in this figure represents r.m.s. fluctuations measured by $\Delta \sigma_{p_t:n}$ (STAR). Corresponding $\Phi_{p_t}$ (CERES) values are numerically the same within 1.5\% (less than the relative errors) within the range of the CERES data~\cite{ceres}, and are therefore plotted without correction on the same scale. 
 
The fluctuation amplitudes in Fig.~\ref{fig3} (right panel) vary strongly with collision centrality (consistent with~\cite{meanptprl}) and energy. The observed trends are fully consistent with $p_t$ angular autocorrelations  reported in~\cite{ptscale} (specifically the near-side minijet peak amplitude). Although there is a trend of monotonic increase with energy for the more central collisions, there is an interesting saturation of $\sqrt{s_{NN}}$ dependence for peripheral collisions ($\nu \leq$ 2.5), consistent with the similarity between left panels in Fig.~\ref{fig2}. 
Also included in this panel is a summary (lower-right hatched box) of $\Phi_{p_t}$ measurements for Pb-Au collisions at 12.3 and 17.3 GeV extrapolated to the STAR $\eta$ acceptance (see left panel) for comparison. 
The lower-left hatched box at $\nu  =1$ represents a $\Phi_{p_t}$ measurement of $2.2\pm 1.5~{\rm (sys)}$ MeV/c for p-p collisions at 17.3 GeV (in the forward rapidity acceptance $y_{\pi} \in [1.1,2.6])$~\cite{na492}. 
We attribute no special significance to the apparent gap between 19.6 and 62.4 GeV data in Fig. 3. In Fig.~\ref{fig4} the data are consistent with the simple logarithmic trend $\ln(\sqrt{s_{NN}})$. The gaps in Fig.~\ref{fig3} arise from a conspiracy of currently available energies and the logarithmic trend.

\section{Errors}


Statistical errors for $\langle p_t \rangle$ fluctuation data in Fig.~\ref{fig1} are 0.005, 0.001, 0.003 and 0.0015 (GeV/c)$^2$ respectively for 19.6, 62.4, 130 and 200 GeV (statistical variations at different scale points are correlated by the integral nature of fluctuation scale dependence). Errors for fluctuation measurements in Fig.~\ref{fig3} (right panel) are indicated by error bars. The upper hatched band in Fig.~\ref{fig3} (right panel) estimates the uncertainty in $\nu$ for 130 GeV data and provides an upper limit for 62.4 and 200 GeV $\nu$ uncertainties (more typically $<3$\%). The uncertainty in $\nu$ for the 20 GeV data is approximately 0.3, as shown by horizontal errors in Fig.~\ref{fig3} (right panel). Systematic corrections to fluctuation amplitudes for tracking inefficiency and backgrounds vary over 15-22\% and 21-35\% ranges for 130 and 200 GeV respectively.
The overall systematic uncertainty for corrected fluctuation amplitudes is 14\%. Autocorrelation errors have two components: statistical fluctuations which survive smoothing and systematic error due to smoothing distortion. Statistical errors for the autocorrelations, estimated by inverting the error estimate for $\Delta \sigma^2_{p_t:n}$, are less than 0.0001 and 0.0003 (GeV/c)$^2$ for autocorrelations at 62.4 and 130 GeV respectively (bin errors are correlated). Smoothing distortions, estimated by passing data through inversion twice,
are less than 5\% of the range of autocorrelation values in each panel.

\section{Energy Dependence}

\begin{figure}[h]
\includegraphics[width=1.65in,height=1.65in]{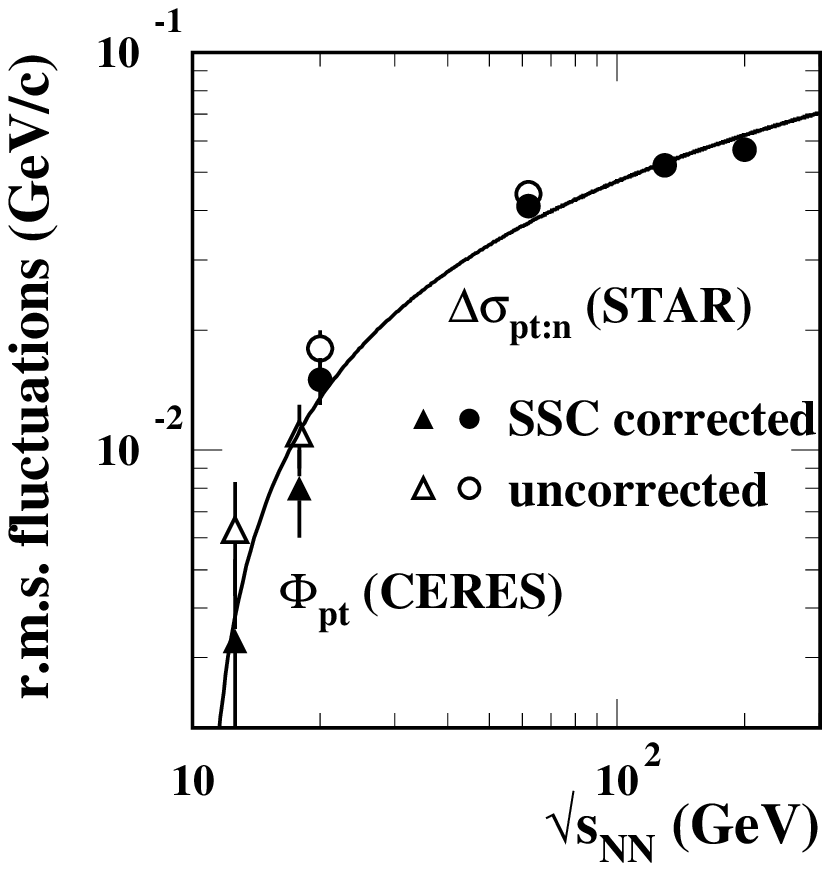}
\includegraphics[width=1.65in,height=1.65in]{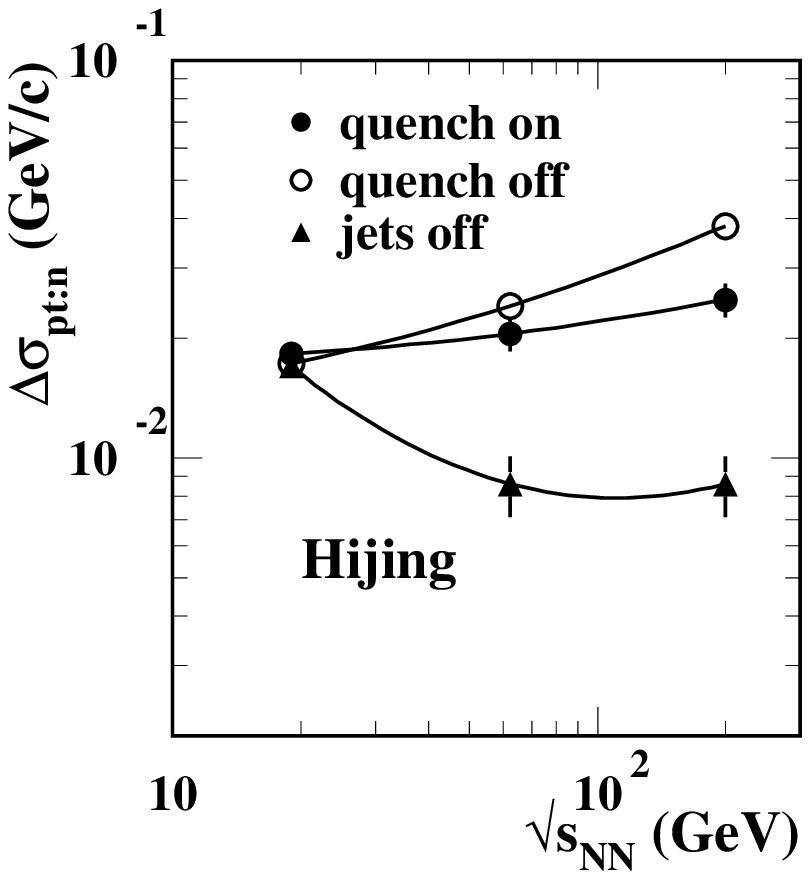}
\caption{Left panel: $\sqrt{s_{NN}}$ dependence of $\langle p_t \rangle$ fluctuations for central collisions and STAR full acceptance. CERES fluctuation data (12.3 and 17.3 GeV) were linearly extrapolated to the STAR $\eta$ acceptance ({\em cf}\, Fig.~\ref{fig3}). The curve is proportional to $\ln\{\sqrt{s_{NN}}/10\}$. Right panel: Hijing-1.37 $\langle p_t \rangle$ fluctuations for central collisions and the STAR acceptance at three energies, for quench-on, quench-off and jets-off collisions.
\label{fig4}}
\end{figure}

As noted previously, although the energy dependence of $p_t$ angular autocorrelations has been established for STAR data by inversion of $\langle p_t \rangle$ fluctuations ({\em e.g.,} Fig.~\ref{fig2} and~\cite{ptscale}) $p_t$ autocorrelations have not been measured at SPS energies. Therefore, we infer the energy dependence indirectly using $\langle p_t \rangle$ fluctuation measurements as proxies. The energy dependence of $\langle p_t \rangle$ fluctuations for STAR and CERES data is summarized in Fig.~\ref{fig4}. Fluctuations measured by the most-central, full-acceptance STAR $\Delta \sigma_{p_t:n}$, and CERES $\Phi_{p_t}$ values linearly extrapolated to $\delta \eta = 2$ (CERES data are linear on $\delta \eta$ in [0.3,0.8], {\em cf.} Fig.~\ref{fig3} -- left panel), are plotted in the left panel {\em vs} $\sqrt{s_{NN}} $, with (solid points) and without (open points) SSC correction. As linear extrapolations the CERES points are actually upper limits, since an NA49 measurement at 17.3 GeV in the pion rapidity interval [1.1,2.6] in the CM gave an upper limit for $\Phi_{p_t}$ of 1.6 MeV/c for central collisions~\cite{na49ptfluct}. A subsequent measurement in the same rapidity interval revealed nonzero results for  more peripheral collisions~\cite{na492}. The SSC correction consists of subtracting 0.003 GeV/c from CERES $\Phi_{p_t}$ and STAR $\Delta \sigma_{p_t:n}$ values (the contribution to the integral of Eq.~(\ref{inverse}) for $\delta \eta < 0.2$). 
Given the uncertainly in the SSC correction and the CERES extrapolations the open points at and below 19.6 GeV represent {upper limits} on a  minijet contribution. The solid curve, proportional to $\ln\{ \sqrt{s_{NN}}/10\}$, summarizes the trend of the data above 10 GeV. We observe a dramatic increase of $\langle p_t \rangle$ fluctuations with beam energy for corrected {\em and} uncorrected data from an apparent onset of minijet-related correlation structure near 10 GeV. 

The fluctuation measure $\Sigma_{p_t} = \sqrt{\Delta\sigma^2_{p_t:n} /\bar{n} \hat{p}_t^2}$ (quantities in the radicand are as defined in this paper) was employed in~\cite{ceres} to suggest that the energy dependence of $\langle p_t \rangle$ fluctuations is negligible, dominated by global temperature fluctuations with $\sigma_T / T_0 \sim 1$\%, unchanging from SPS to RHIC~\cite{ceres,cerqm04}. $\Sigma_{p_t}$ is based on four assumptions: 1) each collision event is thermalized with temperature $T$, 2) an event ensemble has nonzero temperature variance $\sigma^2_T$ about ensemble mean $T_0$, 3) $\Delta \sigma^2_{p_t:n} / \bar n$ estimates $\sigma^2_T$ and 4) $\hat p_{t}$ estimates $T_0$. $\Sigma_{p_t}$ should therefore estimate $\sigma_T / T_0$. All four of those assumptions are falsified by the results reported in this paper. The event-wise thermalization scenario implied by assumptions 1) and 2) is falsified by the strong minijet $p_t$ correlations shown in Fig.~\ref{fig2} and in more detail in~\cite{ptscale}. The observed $p_t$ correlations are almost {\em three times larger} than those predicted for {\em no} thermalization of jets (quench-off Hijing) in~\cite{qingjun}. The separate elements in $\Sigma_{p_t}$ do not estimate temperature-related quantities as implied by assumptions 3) and 4). We learn from Fig.~\ref{fig2} and more extensive results in~\cite{ptscale} that $\Delta\sigma^2_{p_t:n}$ is dominated by jet correlations when evaluated at $\delta \phi = 2\pi$ (the elliptic flow contribution then integrates to zero). The issue of parton scattering and incomplete equilibration is discussed further in Sec.~\ref{discuss}. 

Each of $\bar n$, $\Delta\sigma^2_{p_t:n}$ and $\hat p_t$ varies {\em strongly} with collision centrality and energy. Measured angular $p_t$ autocorrelations combined with two-component trends for $\bar n$ and $\hat p_t$ strongly suggest that those variations are dominated by incompletely-equilibrated semi-hard parton scattering. If the observed hard-scattering contribution were more strongly equilibrated $\Sigma_{p_t}$ might even fall sharply with increasing energy. The algebraic combination $\Sigma_{p_t}$ as measured happens to nearly {cancel} the $\sqrt{s_{NN}}$-dependent hard-scattering trends of the individual factors, but the significance is unclear. A clearer picture emerges when the quantities are studied separately, as in the present study. $\Sigma_{p_t}$ is dominated by reference factor $1/\sqrt{N_{part}}$ and mixes soft (HBT, Coulomb) and hard (parton fragment) contributions (SSC and LSC components respectively) through a {\em running average} on scale $\delta \eta$. $\Sigma_{p_t}$ is thus {by construction} insensitive to the observed energy dependence of parton scattering and fragmentation which dominates $\langle p_t \rangle$ fluctuations at RHIC and which motivated this study.


An analysis using a variant of $\Sigma_{p_t}$ denoted $\sqrt{\langle \delta p_{t,i}\cdot \delta p_{t,j}\rangle}/\langle \langle p_t\rangle \rangle$ was reported in~\cite{ebyepaper}. However, whereas CERES' $\Sigma_{p_t}$ is constructed with the same underlying statistical quantity $\Delta \sigma^2_{p_t:n}$ used in the present analysis, and is therefore directly comparable, the quantity $\sqrt{\langle \delta p_{t,i}\cdot \delta p_{t,j}\rangle}/\langle \langle p_t\rangle \rangle$ is not. In particular, $\langle \delta p_{t,i}\cdot \delta p_{t,j}\rangle$ includes in its denominator the random variable $n(n-1)$ (particle multiplicity $n$ varies randomly from event to event within some limits)~\cite{meanptprl}. For that reason it can produce results significantly inconsistent with $\Phi_{p_t}$ and $\Delta \sigma^2_{p_t:n}$ when bin multiplicities are small. 


For example, in Fig.~\ref{fig2} (lower-left panel) we show results for 80-90\% central Au-Au collisions. The corresponding structure for the Hijing Monte Carlo is very similar: a same-side minijet peak and an away-side ridge from back-to-back jets~\cite{qingjun}. However, the minijet peak from $\bar n \, \langle \delta p_{t,i}\cdot \delta p_{t,j}\rangle$ applied to the {\em same} Hijing data is about $3\times$ smaller, and there is additional (mainly negative) structure of comparable magnitude not present in the  $\Delta \sigma^2_{p_t:n}$ result. Deviations in Hijing data become significant for centralities more peripheral than 50\%. Thus, $\langle \delta p_{t,i}\cdot \delta p_{t,j}\rangle$ cannot be used to study the scale dependence of $p_t$ fluctuations, where the average multiplicity in a bin can be as small as 1.

Analysis of $\langle p_t \rangle$ fluctuations for Hijing-1.37 central collisions in the STAR acceptance at 200 GeV~\cite{QT,qingjun} was extended down to 62.4 and 19.6 GeV for this energy-dependence study. Values of $\Delta \sigma_{p_t:n}$ for quench-on, quench-off and jets-off central Hijing collisions (jet quenching modeled by pQCD gluon bremsstrahlung) are shown in Fig.~\ref{fig4} (right panel). The variation with energy of Hijing $\langle p_t \rangle$ fluctuations is small for jets-on (quench-off and quench-on) collisions. However, the corresponding Hijing $p_t$ autocorrelations ({\em cf.}~\cite{qingjun} for examples at 200 GeV) reveal that the quench-off, near-side minijet peak amplitude falls by about 10$\times$ from 200 to 19.6 GeV, consistent with the variation of  $\langle p_t \rangle$ fluctuations from data shown in the left panel. The source of the inconsistency is apparently the Hijing string fragmentation model. $p_t$ correlations in Hijing below $p_t = 0.5$ GeV/c from {string fragmentation} persist even in central Au-Au collisions and dominate $\langle p_t \rangle$ fluctuations in Hijing at lower energies, consistent with the jets-off results in the right panel. In contrast to Hijing we observe that string-related correlations in RHIC data are rapidly eliminated with increasing Au-Au centrality, even for fairly peripheral collisions~\cite{axialci}.

\section{Discussion} \label{discuss}


Copious minijet production (a minijet `plasma'~\cite{theor2}) has been predicted for Au-Au collisions at RHIC~\cite{theor0,theor1,theor2,theor3,theor4}. The theoretical concept of minijets is low-$Q^2$ partons (mainly gluons) produced in the initial stages of relativistic nuclear collisions~\cite{theor1}. The theoretical parton/minijet $Q$ range extends from a `saturation' limit $Q_s \sim 1$ GeV for RHIC collisions up to several GeV (semi-hard)~\cite{theor2}. Minijets are said to carry most of the transverse energy in central A-A collisions at RHIC~\cite{theor3} and were believed (prior to these measurements) to equilibrate rapidly, driving experimentally-observable hydrodynamic phenomena ({\em e.g.,} elliptic flow)~\cite{theor2}. 
Given the theoretical uncertainty and expected dominance of minijets in the early stages of RHIC collisions it is important to test the observability and degree of equilibration of low-$Q^2$ ($Q \sim$ 1-5 GeV) partons. 

Is a minijet interpretation {\em allowed} for these $p_t$ correlations? A perturbative model of minijet production in heavy ion collisions is stated to apply only above {\em parton} $p_t \sim 2$ GeV/c~\cite{hijing}. However, the authors also state that the 2 GeV/c lower limit is only a limit on the theoretical description, not the physical phenomenon. And, fragments from 2 GeV/c partons should appear at and below 1 GeV/c hadron momentum. p-\=p fragmentation functions for parton energies up to 600 GeV measured at FNAL extend down to 0.35 GeV/c~\cite{fnalfrag}, and e$^+$-e$^-$ fragmentation functions extend to much lower momenta~\cite{tasso,eefrag}. The {\em most probable} fragment momentum in either case is 1-2 GeV/c for a wide range of parton energies. Thus, there is no theoretical or observational reason which could preclude significant jet fragment contributions below 2 GeV/c in heavy ion collisions. Conventional methods for measuring jet angular correlations based on a high-$p_t$ `leading particle' are insensitive to partons below about 6 GeV. However, the novel analysis techniques developed for two-particle correlation analysis described in this paper and elsewhere~\cite{axialci,lowq2,ppmeas,mitinv} have moved the threshold for {\em direct observation} of partons {\em via} final-state hadron correlations down to $Q \sim$ 1 GeV, and jet-like structure is indeed observed. 


Is a minijet interpretation in fact {\em necessary} for these $p_t$ correlations? The correlation structure in Fig.~\ref{fig2} is dominated by a same-side peak, an away-side $\eta_\Delta$-independent ridge and a sinusoid. The sinusoid can be interpreted as elliptic flow (marking its first observation as a velocity phenomenon). A same-side peak (jet cone) and away-side ridge are the expected {\em signature} angular or number correlations for high-$p_t$ hadron fragments from hard-scattered partons. In~\cite{highpt} it was argued that high-$p_t$ angular correlations on azimuth $\phi$ obtained with the leading-particle method in A-A collisions are similar to those obtained in elementary p-p and e$^+$\!- \!e$^-$ collisions with full jet reconstruction and attributed to hard parton scattering. In this analysis the same structure is observed in $p_t$ correlations for $p_t < 2$ GeV/c {without} a high-$p_t$ leading particle. Jet structure dominates peripheral collisions in Fig.~\ref{fig2} (left panels), and jet-like structure in central collisions (right panels), although strongly modified, is part of a continuous shape evolution from N-N collisions. 

Minijet structures have been observed as angular {\em number} correlations of low-$p_t$ particles in Au-Au collisions at 130 GeV~\cite{axialci} and p-p collisions at 200 GeV~\cite{ppmeas}. Low-$p_t$ jet-like structures in $p_t$ and number correlations are observed in Hijing Monte Carlo data where the correlation mechanism is {known} to be parton fragmentation~\cite{qingjun}. In fact, the low-$p_t$ jet-like structure that we observe is exactly what is described as minijets by theory~\cite{minijets}. We conclude therefore that the analogous $p_t$ correlations in this analysis {strongly support} a minijet interpretation in which hadron fragments from {minimum-bias} partons (no condition is imposed by the analysis on the underlying parton momentum distribution) are peaked at low $p_t$. Ironically, low-$Q^2$ partons may be more precisely and unambiguously characterized by high-statistics $p_t$-autocorrelation studies than partons studied in conventional high-$p_t$ leading-particle studies with their biased parton momentum spectra and background subtraction issues.


As noted, $\Delta \rho / \sqrt{\rho_{ref}}$ is within a constant ($\epsilon_\eta\, \epsilon_\phi / \sigma^2_{\hat p_t} \sim 0.25$) Pearson's correlation coefficient, a measure of the {\em relative} covariance between fluctuations in pairs of bins. 
$\Delta \rho / \sqrt{\rho_{ref}}$ for {\em number} correlations measures the number of correlated particle pairs {\em per detected particle}. If $\langle p_t \rangle$ fluctuations are dominated by minijet correlations, as our measurements strongly suggest, then increase of  $\Delta \rho / \sqrt{\rho_{ref}}$ means that either there are relatively more minijets (partons) {\em per detected particle} with the same $p_t$ structure and multiplicity, or the average multiplicity and/or total $p_t$ of minijets has increased, or both. That the minijet correlation structure in Fig.~\ref{fig2} is comparable in amplitude to that identified with elliptic flow suggests that the two dynamical processes have comparable importance in heavy ion collisions at RHIC energies. The trend with $\sqrt{s_{NN}}$ in Fig.~\ref{fig4} (left panel) suggests strong increase of minijet-associated $p_t$ production at higher energies, and a limit on {\em detectable} parton production (correlated hadron fragments) at lower energies, with the apparent onset of observable parton fragments near $\sqrt{s_{NN}} =$ 10 GeV. 


We now return to the question of parton scattering, minijets and equilibration in heavy ion collisions. The correlation structure of an evolving physical system can be used to track the equilibration process. Equilibrated or `thermalized' systems exhibit a large range of correlation types and degrees, from a Bose condensate to an ideal gas. What is relevant for study of a particular system evolving from a non-equilibrium initial state toward equilibrium is {\em changes} in its correlation structure. In the present case parton fragment correlations observed in elementary N-N collisions should be modified in heavy ion collisions depending on the extent and nature of the equilibration process. We can compare observed correlations in heavy ion collisions with the expectation for linear superposition of N-N collisions in a Glauber representation of transparent nuclei as a limiting case. In this analysis we show that abundant minijet structure associated with elementary N-N collisions and exhibiting energy dependence consistent with QCD expectations for parton scattering survives to kinetic decoupling in central Au-Au collisions. The structure is {\em strongly modified}, but the amplitude is still very significant compared to the N-N reference. That result suggests that any claims of complete thermalization for RHIC heavy ion collisions should be reconsidered.




\section{Summary}

In conclusion, we report measurements at several collision energies of $p_t$ angular autocorrelations on pseudorapidity and azimuth difference variables inferred by inverting the scale dependence of $\langle p_t \rangle$ fluctuations. We also report the energy dependence of the centrality and scale dependence of $\langle p_t \rangle$ fluctuations to provide comparisons with CERES measurements at SPS energies. The $p_t$ autocorrelation distributions are
interpreted as consisting mainly of minimum-bias parton fragments ({\em i.e.,} dominated by minijets). We examine energy dependence over the broadest possible interval by comparing STAR measurements to compatible CERES measurements at the CERN SPS. The results are consistent across energies and experiments: 
1) excess $\langle p_t \rangle$ fluctuations increase fourfold from 20 to 200 GeV for central Au-Au collisions, but saturate above 62 GeV for peripheral Au-Au and p-p collisions; 2) those fluctuations correspond mainly to $p_t$ angular correlations identified as minijets, which in central Au-Au collisions are strongly deformed relative to p-p collisions ({\em e.g.,} the same-side peak is broadened on $\eta_\Delta$); 3) $\langle p_t \rangle$ fluctuations thereby associated with initial-state parton scattering increase with energy proportional to $\ln\{\sqrt{s_{NN}}\}$ above an {onset} of detectable parton fragments near $\sqrt{s_{NN}} =$ 10 GeV. $p_t$ autocorrelations thus reveal substantial parton fragment correlations surviving from initial-state scattering even in central Au-Au collisions. Claims of complete thermalization in central HI collisions should be reexamined in light of these results. The strong energy dependence observed in these correlation and fluctuation data should motivate additional measurements at lower RHIC energies, spaced in energy according to the observed logarithmic trend, to investigate the onset of parton fragment production and the relationship of low-$Q^2$ parton scattering to the QCD phase boundary.


We thank the RHIC Operations Group and RCF at BNL, and the
NERSC Center at LBNL for their support. This work was supported
in part by the Offices of NP and HEP within the U.S. DOE Office 
of Science; the U.S. NSF; the BMBF of Germany; IN2P3, RA, RPL, and
EMN of France; EPSRC of the United Kingdom; FAPESP of Brazil;
the Russian Ministry of Science and Technology; the Ministry of
Education and the NNSFC of China; IRP and GA of the Czech Republic,
FOM of the Netherlands, DAE, DST, and CSIR of the Government
of India; Swiss NSF; the Polish State Committee for Scientific 
Research; SRDA of Slovakia, and the Korea Sci. \& Eng. Foundation.


\begin{thebibliography}{9}
\bibitem{diffvar} Difference variable $x_\Delta = x_1 - x_2$ represents a diagonal coordinate axis of 2D space $(x_1,x_2)$, the other diagonal being sum variable $x_\Sigma = x_1 + x_2$. Coordinate axis $x_\Delta$ is differentiated from interval $\Delta x = x_1 - x_2$ on 1D space $x$.
\bibitem{nu}
Quantity $\nu \approx 2N_{bin} / N_{part}$, where $N_{bin}$ is the average number of N-N collisions experienced by a participant nucleon and $N_{part}$ is the average participant nucleon number,
estimates the mean participant path length in number of encountered nucleons.
\end{thebibliography}
\end{document}